\documentstyle[twoside,fleqn,npb]{article}

\newcommand{\g}{{\sl g}}

\newcommand{\cD}{{\cal D}}

\newcommand{\cV}{{\cal V}}

\newcommand{\cO}{{\cal O}}

\newcommand{\cP}{{\cal P}}
\newcommand{\cN}{{\cal N}}
\newcommand{\cQ}{{\cal Q}}

\newcommand{\cU}{{\cal U}}
\newcommand{\cS}{{\cal S}}
\newcommand{\OO}{\mathop{\otimes}}
\font\cmss=cmss12 
\def\1{\hbox{{1}\kern-.25em\hbox{l}}}
\def\bfZ{\relax{\hbox{\cmss Z\kern-.4em Z}}}

\newcommand{\AmS}{{\protect\the\textfont2
  A\kern-.1667em\lower.5ex\hbox{M}\kern-.125emS}}

\begin{document}

\begin{titlepage}

\centerline{\large \bf $\cN = 1$ supersymmetric constraints
                       for evolution kernels.}

\vspace{10mm}

\centerline{\bf A.V. Belitsky\footnote{Alexander von Humboldt Fellow.},
                D. M\"uller}

\vspace{10mm}

\centerline{\it Institut f\"ur Theoretische Physik, Universit\"at
                Regensburg}
\centerline{\it D-93040 Regensburg, Germany}

\vspace{30mm}

\centerline{\bf Abstract}

\vspace{0.8cm}

We provide a complete set of supersymmetric constraints for the anomalous
dimensions of the conformal twist-two operators to all orders of
perturbation theory. Employing them we derive new relations between the
exclusive evolution kernels and apply them in QCD in order to get
definite predictions at leading order and beyond.

\vspace{30mm}

\centerline{\it Talk given at the}
\centerline{\it 7th International Workshop on Deep Inelastic
                Scattring and QCD}
\centerline{\it DESY-Zeuthen, April 19-23, 1999}

\end{titlepage}

\title{$\cN = 1$ supersymmetric constraints for evolution kernels}

\author{A.V. Belitsky\thanks{Alexander von Humboldt Fellow.}
        and D. M\"uller\address{Institut f\"ur Theoretische Physik,
	                        Universit\"at Regensburg, \\
                                D-93040 Regensburg, Germany}}

\begin{abstract}

We provide a complete set of supersymmetric constraints for the anomalous
dimensions of the conformal twist-two operators to all orders of
perturbation theory. Employing them we derive new relations between the
exclusive evolution kernels and apply them in QCD in order to get
definite predictions at leading order and beyond.

\end{abstract}

\maketitle

In the early days of QCD, Dokshitzer had found a simple relation
between the partonic splitting functions \cite{Dok77} once all
colour factors are identified. About a decade after, six leading
order (LO) constraints in the chiral-even sector and a LO relation
in the chiral-odd sector have been derived making use of the conformal
invariance of the classical $\cN = 1$ super Yang-Mills theory
\cite{BukFroKurLip85}. These equations provide a good testing ground
for available QCD results and, e.g.\ the Dokshitzer relation had served
as a consistency check for the non-trivial two-loop calculation of the
DGLAP kernels in the unpolarized \cite{FurPet80} and polarized
\cite{MerNee96} cases as well as for transversity \cite{Vog98}. However,
until recently the status of the remaining constraints was unclear
\cite{Blu98} and, thus, deserves further study.

To start with, we define the so-called conformal composite operators
in the $\cN = 1$ super Yang-Mills theory in the chiral-even sector:
\begin{eqnarray}
\label{treeCO}
\left\{\!\!\!
\begin{array}{c}
{^Q\! \cO^V} \\
{^Q\! \cO^A}
\end{array}
\!\!\!\right\}_{jl}
\!\!\!\!\!\!\!\!&=&\!\!\!\!\! \frac{1}{2}
\bar\psi^a_+ (i \partial_+)^l\!\!
\left\{\!\!\!
\begin{array}{c}
\gamma_+ \\
\gamma_+ \gamma_5
\end{array}
\!\!\!\right\}
\!C^{3/2}_j\!
\left( \frac{\stackrel{\leftrightarrow}{\cD}_+}{\partial_+} \right)
\!\psi^a_+ , \\
\left\{\!\!\!
\begin{array}{c}
{^G\! \cO^V} \\
{^G\! \cO^A}
\end{array}
\!\!\!\right\}_{jl}
\!\!\!\!\!\!\!\!&=&\!\!\!\!\!
G^{a \perp}_{+ \mu} (i \partial_+)^{l-1}\!\!
\left\{\!\!\!
\begin{array}{c}
g^\perp_{\mu\nu} \\
i\epsilon^\perp_{\mu\nu}
\end{array}
\!\!\!\right\}
\!C^{5/2}_{j - 1}\!
\left(
\frac{\stackrel{\leftrightarrow}{\cD}_+}{\partial_+}
\right)
\!G^{a \perp}_{\nu +}, \nonumber\\
\left\{\!\!\!
\begin{array}{c}
\cV \\
\cU
\end{array}
\!\!\!\right\}_{jl}
\!\!\!\!&=&\!\!\!\!\!
\frac{(j + 2)(j + 3)}{(j + 1)} \nonumber\\
&\times&\!\!\!\! G^{a \perp}_{+ \mu} (i \partial_+)^l
P^{(2,1)}_j\!\! \left(
\frac{\stackrel{\leftrightarrow}{\cD}_+}{\partial_+}
\right)
\gamma^\perp_\mu
\left\{\!\!\!
\begin{array}{c}
1 \\
\gamma_5
\end{array}
\!\!\!\right\}
\psi^a_+ , \nonumber
\end{eqnarray}
where $\epsilon^\perp_{\mu\nu} \equiv \epsilon_{\mu\nu-+}$,
$\stackrel{\phantom{\rightarrow}}{\partial}
= \stackrel{\rightarrow}{\partial} \!\!+\!\!
\stackrel{\leftarrow}{\partial}$ and $\stackrel{\leftrightarrow}{\cD}
\, = \stackrel{\rightarrow}{\cD}\! -\! \stackrel{\leftarrow}{\cD}$.
These operators transform covariantly under supersymmetric
transformations and form the irreducible Wess-Zumino chiral
supermultiplet (here $\sigma_j = [1 - (-1)^j]/2$)
\begin{eqnarray}
\label{SUSYtransfS}
&&\!\!\!\!\!\!\!\!\!\!\!\!\delta^Q\, \cS^1_{jl} = \sigma_j\
\bar\zeta \cV_{j - 1 l}, \qquad
\delta^Q\, \cS^2_{jl} = \sigma_j\
\bar\zeta \cV_{jl},  \\
&&\!\!\!\!\!\!\!\!\!\!\!\!\delta^Q\, \cP^1_{jl} = \sigma_{j + 1}\
\bar\zeta \cU_{j - 1 l}, \quad
\delta^Q\, \cP^2_{jl} = \sigma_{j + 1}\
\bar\zeta \cU_{jl}, \nonumber\\
\label{SUSYtransfV}
&&\!\!\!\!\!\!\!\!\!\!\!\!\delta^Q \cV_{j - 1l - 1}
= - \gamma_-\zeta
\left\{
\cS^1_{jl} + \cS^2_{j - 1 l}
\right\} \nonumber\\
&&\qquad\qquad\qquad - \gamma_-\gamma_5\zeta
\left\{
\cP^1_{jl} + \cP^2_{j - 1 l}
\right\} , \nonumber
\end{eqnarray}
where we have introduced particular combinations of the boson operators
\begin{eqnarray}
\label{SUSYmultiplet}
\left\{\!\!\!
\begin{array}{c}
\cS^1 \\
\cP^1
\end{array}
\!\!\!\right\}_{jl}
\!\!\!\!&=&\!\!\!\!
\frac{6}{j}
{^G\! \cO}^{\mit\Gamma}_{jl}
+
{^Q\! \cO}^{\mit\Gamma}_{jl}, \\
\left\{\!\!\!
\begin{array}{c}
\cS^2 \\
\cP^2
\end{array}
\!\!\!\right\}_{jl}
\!\!\!\!&=&\!\!\!\!
\frac{6}{j + 1}
{^G\! \cO}^{\mit\Gamma}_{jl}
-
\frac{j + 3}{j + 1}
{^Q\! \cO}^{\mit\Gamma}_{jl}, \nonumber
\end{eqnarray}
with ${\mit\Gamma} = V(A)$ standing for the $\cS$ ($\cP$) operator.

Similarly, we introduce for the chiral-odd sector the following composite
boson and fermion operators, respectively,
\begin{eqnarray}
{^Q\!{\cal O}^T_{\mu;jl}}
\!\!\!\!\!&=&\!\!\!\! \frac{1}{2}
\bar\psi^a_+ (i \partial_+)^l\!
\gamma_+ \gamma_\mu^\perp
C^{3/2}_j\!
\left( \frac{\stackrel{\leftrightarrow}{\cD}_+}{\partial_+} \right)
\!\psi^a_+ , \\
{^G\!{\cal O}^T_{\mu\nu;jl}}
\!\!\!\!\!&=&\!\!\!\!
G^{a \perp}_{+ \rho} (i \partial_+)^{l-1}
\tau^\perp_{\mu\nu;\rho\sigma}
C^{5/2}_{j - 1}\!
\left(
\frac{\stackrel{\leftrightarrow}{\cD}_+}{\partial_+}
\right)
\!G^{a \perp}_{\sigma +}, \nonumber\\
{\mit\Theta}_{\mu\nu;jl}
\!\!\!\!\!&=&\!\!\!\!
(j + 2)(j + 3) \nonumber\\
&\times&\!\!\!\!
\tau^\perp_{\mu\nu;\rho\sigma}
\gamma^\perp_\rho
G^{a \perp}_{\sigma +}
(i \partial_+)^l
P^{(2,1)}_j \!
\left(
\frac{\stackrel{\leftrightarrow}{\cD}_+}{\partial_+}
\right)
\psi^a_+ , \nonumber
\end{eqnarray}
where $\tau^\perp_{\mu\nu;\rho\sigma} \equiv \frac{1}{2}(
g^\perp_{\mu\rho}g^\perp_{\nu\sigma} + g^\perp_{\mu\sigma}
g^\perp_{\nu\rho} - g^\perp_{\mu\nu}g^\perp_{\rho\sigma})$.
They transform under $\cN = 1$ supertransformations as
\begin{eqnarray}
\label{SUSYtranTransverse}
&&\!\!\!\!\!\!\!\!\!\!\!\delta^Q\, {^Q\!{\cal O}^T_{\mu;jl}} \\
&&\!= \frac{\sigma_j}{2j + 3}
\left\{
\bar\zeta \gamma^\perp_\nu {\mit\Theta}_{\mu\nu;j - 1,l}
- \bar\zeta \gamma^\perp_\nu {\mit\Theta}_{\mu\nu;jl}
\right\}, \nonumber\\
&&\!\!\!\!\!\!\!\!\!\!\!\delta^Q\, {^G\!{\cal O}^T_{\mu\nu;jl}}
\nonumber\\
&&\!= \frac{1}{6} \frac{\sigma_j}{2j + 3}
\left\{ (j + 3)\ \bar\zeta {\mit\Theta}_{\mu\nu;j - 1,l}
+ j\ \bar\zeta {\mit\Theta}_{\mu\nu;jl}
\right\} ,\nonumber\\
&&\!\!\!\!\!\!\!\!\!\!\!\delta^Q\, {{\mit\Theta}_{\mu\nu;jl}}
= - 6 \left\{
\tau^\perp_{\mu\nu;\rho\sigma} \gamma_- \zeta
+ i \epsilon^\perp_{\mu\nu;\rho\sigma} \gamma_- \gamma_5 \zeta
\right\} \nonumber\\
&&\!\times\left\{
{^G\!{\cal O}^T_{\mu\nu;j + 1,l + 1}}
+
{^G\!{\cal O}^T_{\mu\nu;j,l + 1}}
\right\} \nonumber\\
&&\!- \frac{1}{2}
\tau^\perp_{\mu\nu;\rho\sigma} \gamma^\perp_\rho
\left\{
g^\perp_{\lambda\sigma} \gamma_- \zeta
+ i \epsilon^\perp_{\lambda\sigma} \gamma_- \gamma_5 \zeta
\right\} \nonumber\\
&&\!\times \left\{
(j + 1)\
{^Q\!{\cal O}^T_{\lambda;j + 1,l + 1}}
- (j + 3)\
{^Q\!{\cal O}^T_{\lambda;j,l + 1}}
\right\}. \nonumber
\end{eqnarray}
Here $\epsilon^\perp_{\mu\nu;\rho\sigma} \equiv \frac{1}{2}
(\epsilon^\perp_{\mu\rho} g^\perp_{\nu\sigma} + \epsilon^\perp_{\nu\rho}
g^\perp_{\mu\sigma} )$.

In order to derive the desired constraints for the anomalous dimensions,
we use the commutator relation between the generators of dilatation $\cD$
and supersymmetric transformation $\cQ$:
\begin{eqnarray}
\label{commutator}
\left[ \cQ, \cD \right]_- = \frac{i}{2} \cQ .
\end{eqnarray}
The infinitesimal dilatation, $\cD$, is equivalent to a change in
the scale, and, therefore, it is governed by the renormalization
group equation
\begin{eqnarray}
\label{RGE}
\mu \frac{d}{d\mu} [\mbox{\boldmath${\cal O}$}_{jl}]
= - \sum_{k = 0}^{j} \mbox{\boldmath${\gamma}$}^{\cal O}_{jk}
[\mbox{\boldmath${\cal O}$}_{kl}],
\end{eqnarray}
with anomalous dimensions $\mbox{\boldmath${\gamma}$}^{\cal O}
= \{\lambda_{jk},\, \mbox{\rm for}\, {\cal O} = {\cal U},{\cal V};\
\omega_{jk},\, \mbox{\rm for}\, {\cal O} = {\mit\Theta} \}$.
Applying now the commutator (\ref{commutator}) on the Green functions
with conformal operators, $\langle [\mbox{\boldmath${\cal O}$}_{jl}]
\prod_i \phi (x_i) \rangle$, and making use of Eqs.\ (\ref{SUSYtransfS})
and (\ref{RGE}, we derive the following constraints for eight
anomalous dimensions in the chiral-even sector
\begin{eqnarray}
\label{constr-even}
{^{11}\!\gamma}^{\cal S}_{2n + 1, 2m + 1}
\!\!\!&=&\!\!\! {^{22}\!\gamma}^{\cal P}_{2n, 2m}
= \lambda_{2n, 2m} , \\
{^{12}\!\gamma}^{\cal S}_{2n + 1, 2m + 1}
\!\!\!&=&\!\!\! {^{21}\!\gamma}^{\cal P}_{2n, 2m + 2}
= \lambda_{2n, 2m + 1} , \nonumber\\
{^{21}\!\gamma}^{\cal S}_{2n + 1, 2m + 1}
\!\!\!&=&\!\!\! {^{12}\!\gamma}^{\cal P}_{2n + 2, 2m}
= \lambda_{2n + 1, 2m} , \nonumber\\
{^{22}\!\gamma}^{\cal S}_{2n + 1, 2m + 1}
\!\!\!&=&\!\!\! {^{11}\!\gamma}^{\cal P}_{2n + 2, 2m + 2}
= \lambda_{2n + 1, 2m + 1}.
\nonumber
\end{eqnarray}
Since we did not employ the conformal symmetry, which is obviously
broken, we obtain the true constraints where both the diagonal
(conformal) and the non-diagonal (anomalous) matrix elements are
involved. Note that the restriction to the forward case provides
six relations, however, two of them necessarily involve the
non-diagonal elements of the anomalous dimension matrix of
the conformal operators \cite{BelMulSch99}. Omitting this non-diagonal
part inevitably lead to the violation of these constraints beyond LO
\cite{Blu98}.

Analogously, using Eq.\ (\ref{SUSYtranTransverse}) we find two
relations in the chiral-odd sector ($v_n \equiv n + 2$, $w_n \equiv
2 n + 1$)
\begin{eqnarray}
&&\!\!\!\!\!\!\!\!\!\!\!\!
\frac{4n + 5}{4m + 5}{^{QQ}\!\gamma}^{T}_{2n + 1, 2m + 1} \\
&&=
\left\{
\omega_{2n + 1, 2m + 1} - \omega_{2n, 2m + 1}
\right\}, \nonumber\\
&&\!\!\!\!\!\!\!\!\!\!\!\!
\frac{4n + 5}{4m + 5}{^{GG}\!\gamma}^{T}_{2n + 1, 2m + 1}
\nonumber\\
&&=
\left\{
\frac{w_n}{w_m} \omega_{2n + 1, 2m + 1}
+ 2 \frac{v_n}{w_m} \omega_{2n, 2m + 1}
\right\}.
\nonumber
\end{eqnarray}
In the forward case we find then one equation:
${^{QQ}\!\gamma}^{T}_{2n + 1, 2n + 1}
= {^{GG}\!\gamma}^{T}_{2n + 1, 2n + 1}
= \omega_{2n + 1, 2n + 1} $. Note also that there
exist two relations between the anomalous dimensions
$\omega_{jk}$ which read
\begin{eqnarray}
&&\!\!\!\!\!\!\!\!\!\!\! \omega_{2n + 1, 2m + 1} - \omega_{2n, 2m + 1} =
\omega_{2n, 2m} - \omega_{2n + 1, 2m}\ ,
\nonumber\\
&&\!\!\!\!\!\!\!\!\!\!\!\! \frac{w_n}{w_m} \omega_{2n + 1, 2m + 1}
+ 2 \frac{v_n}{w_m} \omega_{2n, 2m + 1}
\\
&& =\frac{v_n}{v_m} \omega_{2n, 2m}
+ \frac{1}{2} \frac{w_n}{v_m} \omega_{2n + 1, 2m}.
\nonumber
\end{eqnarray}

Let us address the question of use of these equations in QCD. Instead of
direct calculations of otherwise different quark-gluon anomalous dimensions
at LO, when conformal covariance holds true, we can obtain them from the
one-loop anomalous dimensions of fermion operators in $\cN = 1$ super
Yang-Mills theory ($\sigma (j) \equiv (- 1)^j$):
\begin{eqnarray}
\frac{\lambda_j}{N_c}
\!\!\!\!\!&=&\!\!\!\!
2 \psi (j + 1)\! + 2 \psi (j + 4)\! - 4 \psi (1) \\
&-&\!\!\!\!\! \frac{4 \sigma (j)}{(j + 1 )(j + 2)(j + 3)} - 3 ,
\nonumber\\
\frac{\omega_j}{N_c}
\!\!\!\!\!&=&\!\!\!\!
2 \psi (j + 2)\! + 2 \psi (j + 3)\! - 4 \psi (1)
\!+\! \frac{2 \sigma (j)}{j + 2} - 3 . \nonumber
\end{eqnarray}
Obviously, at LO the colour structure, arising in QCD, can be restored in
a unique way.

The results for the anomalous dimensions (\ref{constr-even}) imply
constraints for the evolution kernels. While for the splitting
functions they are known since long ago, we present them now also
for the ER-BL evolution kernels. As demonstrated below, these constraints
are also useful beyond LO to reconstruct conformal parts of the exclusive
kernels. Since at LO conformal covariance holds true, the ER-BL evolution
kernels are given as a single sum over Gegenbauer polynomials
\cite{BelMul98c} and allow to derive the following relations
$(\bar x \equiv 1 - x)$:
\begin{eqnarray}
&&\!\!\!\!\!\!\!\!\!\!\!\!\frac{d}{dy}
{^{QQ} V^i} (x, y) + \frac{d}{dx} {^{GG} V^i} (x, y)
= - 3 {^{QG} V^i} (x, y), \nonumber\\
&&\!\!\!\!\!\!\!\!\!\!\!\!{^{GQ} V^i} (x, y)
= \frac{(\bar x x)^2}{\bar y y} {^{QG} V^i} (y, x),
\end{eqnarray}
for $i = \{ V, A \}$. There also exist two differential equations
relating the parity odd with the parity even sectors. Thus, from
the knowledge of the $QQ$-channel, we find all other kernels by
solving six differential equations and requiring their solutions
to respect conformal covariance. Thus, we predict unambiguously
the results for the mixed LO kernels ${^{AB} V^i} = \theta(y - x)
{^{AB} F^i} (x, y) - \left\{x \to \bar x \atop y \to \bar y \right\}$:
\begin{eqnarray}
{^{QG} F^i}\!\!\!\! &=&\!\!\!\! 2 T_F N_f \frac{x}{y^2}
\left\{\!\!\!
\begin{array}{ll}
- 1 + 2 x - 4 \bar x y, & \!\!\!\mbox{for}\ i = V \\
- 1,                    & \!\!\!\mbox{for}\ i = A
\end{array}
\right. \!\!\!\!, \nonumber\\
{^{GQ} F^i}\!\!\!\! &=&\!\!\!\! C_F \frac{x^2}{y}
\left\{\!\!\!
\begin{array}{ll}
- 1 + 2 y - 4 \bar x y, & \!\!\!\mbox{for}\ i = V \\
\phantom{-} 1,          & \!\!\!\mbox{for}\ i = A
\end{array}
\right. \!\!\!\! ,
\end{eqnarray}
which, therefore, resolve the confusion about diversity of
kernels available in the literature.

Now we address a more interesting issue of applying supersymmetric
constraints beyond LO. Let us first mention that in the chiral-even
sector the NLO forward anomalous dimensions and the NLO non-diagonal
entries, derived from conformal constraints and a one-loop calculation
of the special conformal anomalies, fulfill all superconstraints
(\ref{constr-even}). Of course, they do not hold in the conventional
dimensional regularization (DREG) scheme but rather in a scheme which
preserves supersymmetry (DRED). The transformation
$\mbox{\boldmath$\gamma$}^{\rm DRED} = \mbox{\boldmath$z$}
\mbox{\boldmath$\gamma$}^{\rm DREG} \mbox{\boldmath$z$}^{-1}
- \beta (\g) \frac{\partial}{\partial\g}
\mbox{\boldmath$z$} \cdot \mbox{\boldmath$z$}^{-1}$ is driven by
rotation
\begin{eqnarray*}
\mbox{\boldmath$z$}_{jk}
\!=\! \1 \delta_{jk}\! +\! \frac{\alpha_s}{2 \pi} N_c
\left\{
\mbox{\boldmath$z$}^{\rm D}_j \delta_{jk}
\!+\!
2 \mbox{\boldmath$z$}^{\rm ND}_{jk}
\theta_{j - 2,k} \sigma_{j - k + 1}
\right\} ,\!\!\!\!\!\!\!\!\!\!
\end{eqnarray*}
with the following matrices
\begin{eqnarray}
\label{TransDiag}
&&\!\!\!\!\!\!\!\!\mbox{\boldmath$z$}^{{\rm D},V}_j
= \left(
\begin{array}{cc}
- \frac{j(j + 3)}{2(j + 1)(j + 2)}
&
\frac{12}{j(j + 2)(j + 3)}
\\
\frac{j}{6 (j + 2)}
&
-\frac{1}{6}
\end{array}
\right) , \\
&&\!\!\!\!\!\!\!\!\mbox{\boldmath$z$}^{{\rm D},A}_j
= \left(
\begin{array}{cc}
- \frac{j(j + 3)}{2(j + 1)(j + 2)}
&
\frac{12}{j(j + 1)(j + 2)}
\\
- \frac{j}{3 (j + 1)(j + 2)}
&
-\frac{1}{6} - \frac{4}{(j + 1)(j + 2)}
\end{array}
\right) , \nonumber\\
\label{TransNonDiag}
&&\!\!\!\!\!\!\!\!\mbox{\boldmath$z$}^{{\rm ND},V}_{jk}
= \mbox{\boldmath$z$}^{{\rm ND},A}_{jk} \nonumber\\
&&\!\!\!\!\!\! = \left(
\begin{array}{cc}
0
&
\frac{6 (2k + 3)}{k( k + 1 )( k + 2 )( k + 3 )} \\
- \frac{(2k + 3)}{6( k + 1 )( k + 2 )}
&
- \frac{(2k + 3)(j - k)(j + k + 3)}{k(k + 1)(k + 2)(k + 3)}
\end{array}
\right) . \nonumber
\end{eqnarray}

Moreover, from conformal constraints we have found a simple
representation for the NLO corrections to the ER-BL kernels,
which reads in matrix notation ($\OO \equiv \int_{0}^{1}$)
\begin{eqnarray*}
\mbox{\boldmath $V$}^{(1)}
\!=\! - \!\mbox{\boldmath $\dot{V}$}\!
\OO\! \left(
\mbox{\boldmath $V$}^{(0)}\! + \frac{\beta_0}{2} \1
\right)
-\! \Big[ \mbox{\boldmath $g$}
\OO_{\mbox{'}} \mbox{\boldmath$V$}^{(0)}\! \Big]_-
\!\!\!\! +\! \mbox{\boldmath $D$}\! + \! \mbox{\boldmath $G$} ,
\!\!\!\!\!\!\!\!\!\!\!\!\!\!
\end{eqnarray*}
where the {\boldmath $g$}, {\boldmath $\dot{V}$}, and
$\mbox{\boldmath $V$}^{(0)}$ are known \cite{BelMul98c}. The diagonal
kernel {\boldmath $G$} is defined by contributions coming from the
two-loop crossed ladder diagram (in the light-cone gauge) and is known
in the $QQ$-channel. Since this diagram does not contain divergent
subgraphs, it respects supersymmetry and tree level conformal
covariance. From this information, similar to LO case, one can construct
the {\boldmath $G$} kernels in all other channels. The remaining diagonal
piece, $\mbox{\boldmath $D$}$, can be represented as convolutions of
simple kernels. Proceeding along this line we reconstruct the whole
ER-BL kernel in NLO without explicit two-loop calculations
\cite{BelMulFre99a}.

{\bf Acknowledgements.}  A.B. was supported by the Alexander von
Humboldt Foundation.

\end{document}